\documentclass[aps,amssymb,amsmath,singlecolumn,pra,superscriptaddress]{revtex4}
\usepackage[latin9]{inputenc}
\usepackage{color}
\usepackage{amsmath}
\usepackage{graphicx}
\usepackage[unicode=true,
 bookmarks=false,
 breaklinks=false,pdfborder={0 0 1},backref=non,colorlinks=false]
 {hyperref}
\hypersetup{
 colorlinks,citecolor=blue}

\makeatletter
\@ifundefined{textcolor}{}
{%
 \definecolor{BLACK}{gray}{0}
 \definecolor{WHITE}{gray}{1}
 \definecolor{RED}{rgb}{1,0,0}
 \definecolor{GREEN}{rgb}{0,1,0}
 \definecolor{BLUE}{rgb}{0,0,1}
 \definecolor{CYAN}{cmyk}{1,0,0,0}
 \definecolor{MAGENTA}{cmyk}{0,1,0,0}
 \definecolor{YELLOW}{cmyk}{0,0,1,0}
}

\makeatother

\begin{document}
\title{Quantum direct communication protocols using discrete-time quantum
walk}
\author{Srikara S}
\affiliation{Indian Institute of Science Education and Research, Pune-411008, India}
\author{C. M. Chandrashekar}
\email{chandru@imsc.res.in}

\affiliation{The Institute of Mathematical Sciences, C. I. T. Campus, Taramani,
Chennai 600113, India}
\affiliation{Homi Bhabha National Institute, Training School Complex, Anushakti
Nagar, Mumbai 400094, India}
\begin{abstract}
The unique features of quantum walk, such as the
possibility of the walker to be in superposition of the position space and get entangled with the position space, provides inherent advantages that can be captured to design highly secure quantum communication
protocols. Here we propose two quantum direct communication protocols,
a Quantum Secure Direct Communication (QSDC) protocol and a Controlled
Quantum Dialogue (CQD) protocol using discrete-time quantum walk on
a cycle. The proposed protocols are unconditionally secure against
various attacks such as the intercept-resend attack, the denial of
service attack, and the man-in-the-middle attack. Additionally, the
proposed CQD protocol is shown to be unconditionally secure against
an untrusted service provider and both the protocols are shown more
secure against the intercept resend attack as compared to the qubit
based LM05/DL04 protocol. 
\end{abstract}
\maketitle

\section{Introduction}


The research in quantum cryptography, which first started with the
BB84 quantum key distribution (QKD) protocol \cite{key-1} was later
followed up with the design and the study of various novel QKD schemes
\cite{key-2,key-3,key-4}. These protocols were designed to securely
generate a secret key between two parties, which would then be used
to encode the message via a one-time pad. Most of the research in
quantum cryptography was concentrated on QKD, until during 2002-05,
when few new protocols were introduced \cite{LL02, key-5, DLL03, DL04, key-6}.  
These protocols were called Quantum Secure Direct Communication (QSDC) protocols. In 2002, the first QSDC
protocol was proposed in the form of a deterministic key that could
transmit the secret message \cite{key-5, LL02}. In 2003, the standard form
of QSDC without the requirement of a key was proposed \cite{DLL03}
and in 2004 and 2005 a single photon based protocol called DL04 was proposed
\cite{DL04}. Recently, the Device-independent QSDC protocol \cite{DI20}
and measurement-device-independent QSDC protocol \cite{MDI18,MDI20}
have also been proposed. In 2004, a two-way quantum direct communication
protocol was introduced, called the quantum dialogue(QD) \cite{key-7}.
Unlike QSDC protocols where communication is just one way, in QD protocols
both the parties interact with each other i.e., communication is two
ways. This quantum dialogue protocol was extended to a controlled
quantum dialogue protocol (CQD), in which a third party provides the
quantum services for communication \cite{key-8}. The QSDC, QD and
the CQD protocols have shown that an unconditionally secure quantum
communication can be achieved even without a key. With the reporting
of experimental realization of QSDC protocols \cite{Expt1,Expt2,Expt3},
its significance for practical use is being highlighted.

In 1993, the concept of quantum walks was introduced \cite{key-9}.
Quantum walks are the quantum analogues of classical random walks.
Unlike classical random walks where the walker is at just one deterministic
position at a given time, in quantum walks, the walker can be at multiple
positions at the same time, i.e., in a superposition of position space.
The tossed quantum coin that decides the movement of the walker, can
also be at a superposition of head and tails. These unique features
of quantum walks can help traverse multiple positions faster, a feature
that has been exploited in the design of various quantum search algorithms
\cite{key-10}. Quantum walks have also been used for studying and
describing various quantum phenomena \cite{key-11,key-12} and also
in the study and design of quantum networks \cite{key-13}. Surprisingly,
the usage of quantum walks for cryptography and secure communication
has largely been unexplored, except for a few designs of QKD protocols
\cite{key-14} and public-key cryptosystems \cite{key-15}. In this
work, we delve into an unexplored cryptographic potential of quantum
walks, which is the quantum direct communication. Using the discrete-time
quantum walk on a cycle, we propose two new protocols for QSDC and
CQD and show the unconditional security they provide against various
attacks such as the intercept-resend attack, the denial of service
attack and the man-in-the-middle attack. We also show that the proposed
CQD protocol provides unconditional security against an untrusted
service provider and both the protocols are more secure against the
intercept resend attack as compared to the qubit based LM05/DL04 protocol \cite{key-6, DL04}.

This paper is structured as follows: In section \ref{sec:Discrete-Time-Quantum},
we introduce the preliminary concepts of discrete-time quantum walk
on a cycle required to understand the protocols proposed in section
\ref{sec:Protocols}. In section \ref{sec:Security}, we discuss the
security of the proposed protocols against various attacks. In section
\ref{sec:Conclusion}, we conclude with our remarks. In the Appendix,
we provide relevant background details that can be referred to if
required.


\section{Discrete-time quantum walk on a cycle - preliminaries\label{sec:Discrete-Time-Quantum}}

Quantum walks are a quantum analogue of the classical random walks.
In discrete-time quantum walk on an $N$-cycle, the walker moves along
$N$ discrete points on a cycle \cite{qwcycle}, which are represented
by $N$-dimensional quantum states $|x\rangle$, orthogonal to each
other and belonging to the Hilbert space $H_{p}$ where 
\begin{center}
\[
H_{p}=span\{|x\rangle,x\in\{0,1,2,...,N-1\}\}.
\]
. 
\par\end{center}

\noindent During each step of the discrete-time quantum walk, the
walker moves one position either to his left or to his right based
on the result ($|0\rangle$ or $|1\rangle$) of the quantum coin,
which is given by a two-dimensional quantum state $|c\rangle$ belonging
to the Hilbert space $H_{c}$ where 
\begin{center}
\[
H_{c}=span\{|0\rangle,|1\rangle\}.
\]
\par\end{center}

If the walker is in a superposition of the coin state, it will move
to both, left and right simultaneously, creating a state which is
in superposition in position space. Thus, the initial state of the
walker starting at position $x_{in}$ and with an initial coin state
$|c_{in}\rangle$ can be considered to be in a superposition of the
two allowed basis states given by 
\begin{center}
\begin{equation}
|\Psi_{in}\rangle=|x_{in}\rangle\otimes|c_{in}\rangle=|x_{in}\rangle|c_{in}\rangle~~~~;~~~~|x_{in}\rangle\in H_{p}~~~~;~~~~|c_{in}\rangle\in H_{c}.
\end{equation}
\par\end{center}

\noindent The dynamics of the walker during each step of the walk
is governed by the action of the unitary operator, a composition of
a quantum coin operation on the coin space followed by a conditioned
position shift operation on the complete Hilbert space \cite{key-16,meyer1996quantum,SU2}, 
\begin{center}
\begin{equation}
U=U(\theta,\xi,\zeta)=S(I_{p}\otimes R_{c}).
\end{equation}
\par\end{center}

\noindent Here $I_{p}$ is the identity operator on position space
and the quantum coin operation $R_{c}$ is given by, 
\begin{center}
\begin{equation}
R_{c}=R_{c}(\theta,\xi,\zeta)=\begin{bmatrix}e^{i\xi}\cos\theta & e^{i\zeta}\sin\theta\\
e^{-i\zeta}\sin\theta & e^{-i\xi}\cos\theta
\end{bmatrix}.
\end{equation}
\par\end{center}

\noindent In simpler cases, when $\zeta=\xi=0$ or fixed to a specific
value, $R_{c}(\theta,\xi,\zeta)=R_{c}(\theta)$ is the coin operator
on the coin space. The shift operator on $H=H_{p}\otimes H_{c}$,
which shifts the position of the walker in the direction which is
determined by the coin state is given by 
\begin{equation}
S=\overset{N-1}{\underset{x=0}{\Sigma}}(|x-1(\mod N)\rangle\langle x|\otimes|0\rangle\langle0|+|x+1(\mod N)\rangle\langle x|\otimes|1\rangle\langle1|).
\end{equation}
The state after $t$ steps of the walk on an $N-$cycle, in general,
will be in the form, 
\begin{equation}
|\Psi_{t}\rangle=U^{t}|\Psi_{in}\rangle=\sum_{x=1}^{N}|x\rangle\otimes\Big(\alpha_{x,t}|0\rangle+\beta_{x,t}|1\rangle\Big),\alpha_{x,t}\,\and\,\beta_{x,t}\in\mathbb{C}\,\forall\,x,t
\end{equation}
and the probability of finding the walker at any position $x$ after
$t$ steps of the walk will be $P(x,t)=|\alpha_{x,t}|^{2}+|\beta_{x,t}|^{2}$.
In addition to the quantum walk evolution operator, we will also define
the translation operator and measurement operator which will be needed
for QSCD and CQD protocols. The translation operator
$T$ defined on the space $H_{p}$ is given as:
\begin{center}
\begin{equation}
T(y)=\underset{x=0}{\overset{N-1}{\Sigma}}|x+y(\mod N)\rangle\langle x|
\end{equation}
\par\end{center}

\noindent and the measurement operator $M$ is defined on the entire
space $H$ in the form given by 
\begin{center}
\[
M=M_{p}\otimes M_{c}
\]
\par\end{center}

\noindent where 
\begin{center}
\begin{equation}
M_{p}=\overset{N-1}{\underset{x=0}{\Sigma}}|x\rangle\langle x|~~~~~~~~\mbox{and}~~~~~~~~M_{c}=\overset{1}{\underset{c=0}{\Sigma}}|c\rangle\langle c|.
\end{equation}
\par\end{center}

\noindent Note that $[T(y),U]=0$ i.e., $T(y)$ and $U$ commute with
each other \cite{key-15}.


\section{The protocols\label{sec:Protocols}}


The extent of the spread of the discrete-time quantum walk in position
space is mainly governed by the parameter $\theta$ in the quantum
coin operation \cite{meyer1996quantum,SU2}. Therefore, in this paper,
we will keep only the coin parameter $\theta$ as a variable parameter
while keeping the parameters $\xi$ and $\zeta$ constant throughout
the protocols. Here we first present the encoding scheme, and then
we present the protocols for QSDC and CQD. The schematic representation
of the protocols for QSDC and CQD is presented in Fig.\,\ref{fig1}
and Fig.\,\ref{fig2}, respectively. In both the
figures, the ``random path switcher'' is a device that switches
the path of the quantum channel to move a particular state into encoding
the message or into checking eavesdropping, similar to using a physical
lever that is used for changing the railway tracks. For example, for
linear and quantum optical implementations of quantum walks \cite{qwimp},
various optical switches \cite{os1,os2,os3} coupled with a quantum
random number generator \cite{SC19} can be used as a random path
switcher.


\subsection{Encoding of the message}


The message $m$ (or a part $m$ of the total message) is encoded
on a discrete-time quantum walk state $|\phi\rangle=\underset{i}{\sum}|x_{i}\rangle|c_{i}\rangle$
by applying the translation operator $T(m)$ on $|\phi\rangle$, resulting
in the state $T(m)\otimes I_{c}|\phi\rangle=\underset{i}{\sum}|x_{i}+m(\mod N)\rangle|c_{i}\rangle$.


\subsection{Discrete-time quantum walk based QSDC protocol\label{subsec:Quantum-Walk-based}}


\begin{center}
\begin{figure}[h]
\begin{centering}
\includegraphics[scale=0.6]{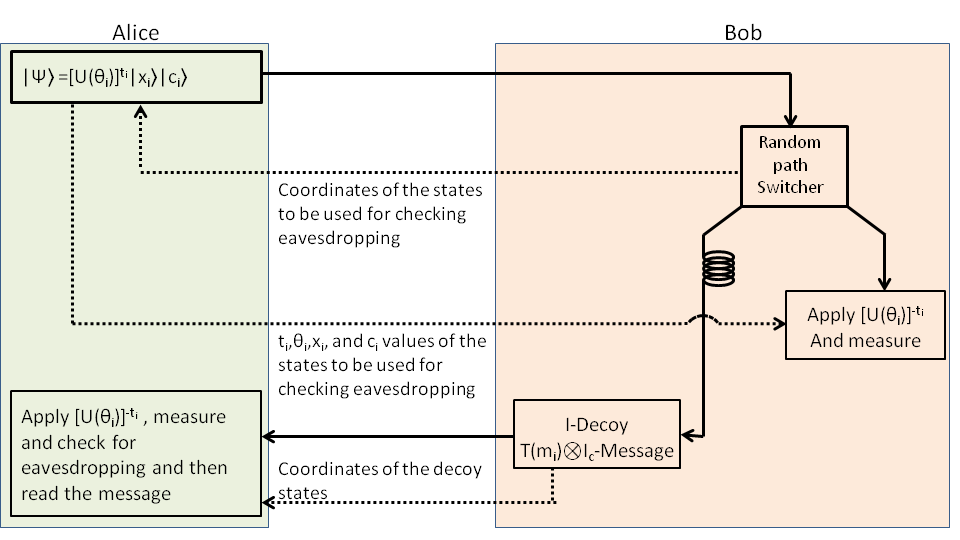} 
\par\end{centering}
\caption{Schematic diagram of the discrete-time quantum walk based QSDC protocol.
The bold arrow lines represent quantum channels whereas the dotted
arrow lines represent classical channels }
\label{fig1} 
\end{figure}
\par\end{center}
\begin{enumerate}
\item Alice prepares $n$ discrete-time quantum walk states. To prepare
$n$ quantum walk states, Alice randomly chooses $3n$ integers \{$t_{1},t_{2},...,t_{n}$\},
\{$x_{1},x_{2},...,x_{n}$\} and \{$c_{1},c_{2},...,c_{n}$\} such
that $x_{i}\in\{0,1,2,...,N-1\}$, $c_{i}\in\{0,1\}$ and $t_{i}\in\mathbb{N}\cup\{0\}$
$\forall i\in\{1,2,...,n\}$ and $n$ random real numbers \{$\theta_{1},\theta_{2},...,\theta_{n}$\}
such that $\theta_{i}\in[0,2\pi]$. Thus, she prepares $n$ discrete-time
quantum walk states $[U(\theta_{i})]^{t_{i}}|x_{i}\rangle|c_{i}\rangle=U^{t_{i}}|x_{i}\rangle|c_{i}\rangle$
~~$\forall i\in\{1,2,...,n\}$ and sends these states to Bob. (In
the rest of this and the next protocol, we will refer to $[U(\theta_{i})]$
as $U$). 
\item On receiving the walk states from Alice, Bob randomly chooses $n/2$
of them for checking eavesdropping and classically sends their corresponding
coordinates $i$ to Alice. Alice classically sends to Bob the corresponding
values of $t_{i},x_{i},c_{i},$ and $\theta_{i}$. Bob applies the
corresponding operation $U^{-t_{i}}$ on those states, measures them,
and checks the measurement result with the value of $x_{i}$ and $c_{i}$.
If the error is within a tolerable limit, he continues to step 3.
Otherwise, the protocol is aborted and they will start the protocol
all over again. 
\item Out of the remaining $n/2$ walk states, Bob chooses $n/4$ of them
for encoding the message. On each of those $n/4$ states, Bob codes
a part of his message $m_{i}$ by applying the translation operator
$T(m_{i})\otimes I_{c}$. He does nothing to the other $n/4$ states
(let us call them decoy states). He then sends all the $n/2$ states
back to Alice. 
\item Once Alice confirms the receiving of the states, Bob classically sends
the coordinates of the decoy states to Alice. Alice applies the corresponding
operator $U^{-t_{i}}$ on the decoy states and checks for eavesdropping
just like how Bob does it in step 2. 
\item Once no eavesdropping is confirmed, Alice then applies $U^{-t_{i}}$
on the remaining $n/4$ message states and measures them to obtain
the message sent by Bob. 
\end{enumerate}

\subsection{Discrete-time quantum walk based CQD protocol\label{subsec:Quantum-Walk-based-1}}


\begin{center}
\begin{figure}[h]
\begin{centering}
\includegraphics[scale=0.4]{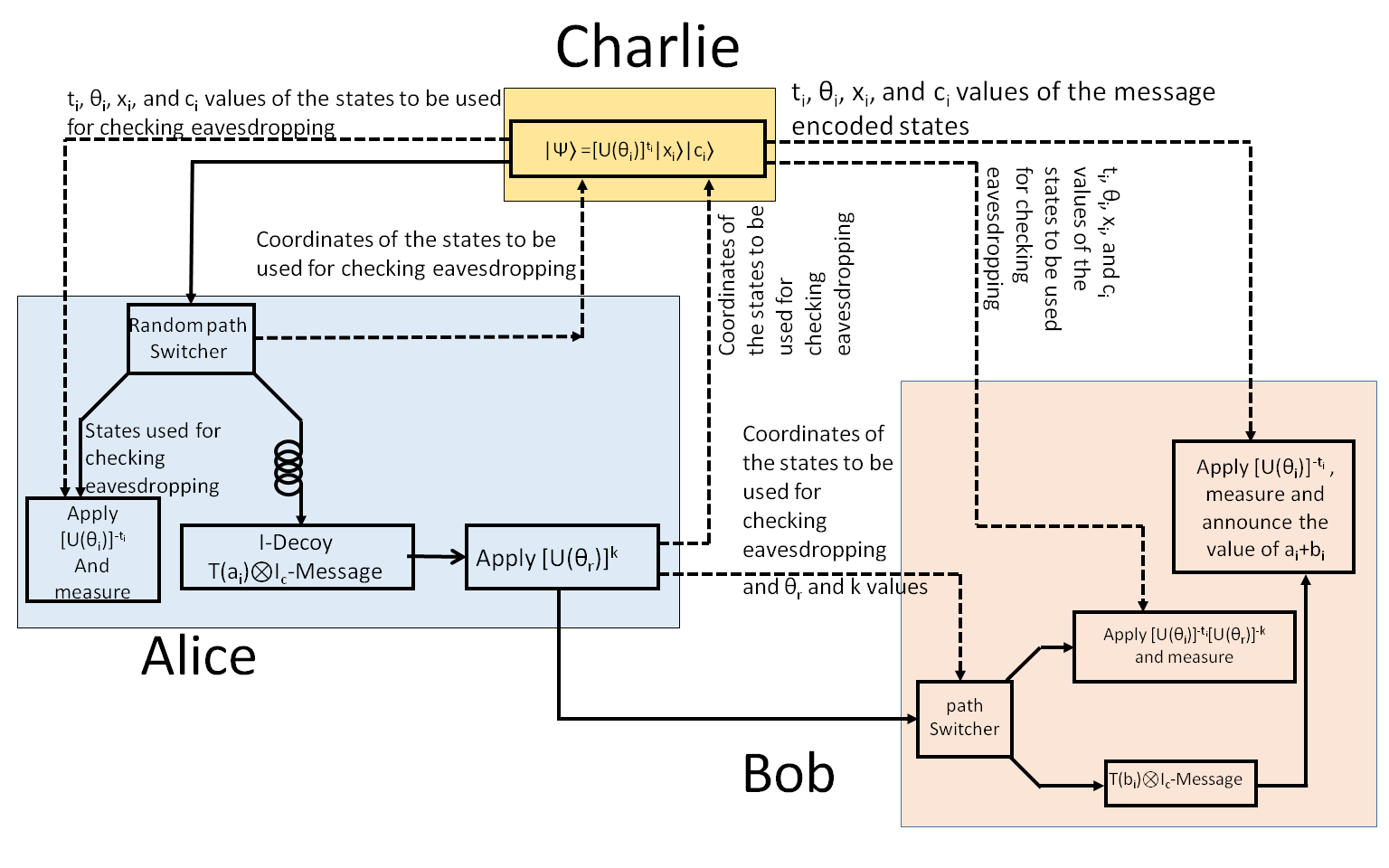} 
\par\end{centering}
\caption{Schematic diagram of the discrete-time quantum walk based CQD protocol.
The bold arrow lines represent quantum channels whereas the dotted
arrow lines represent classical channels.}
\label{fig2} 
\end{figure}
\par\end{center}
\begin{enumerate}
\item Charlie prepares $n$ discrete-time quantum walk states. To prepare
$n$ quantum walk states, Charlie randomly chooses $3n$ integers
\{$t_{1},t_{2},...,t_{n}$\}, \{$x_{1},x_{2},...,x_{n}$\} and \{$c_{1},c_{2},...,c_{n}$\}
such that $x_{i}\in\{0,1,2,...,N-1\}$, $c_{i}\in\{0,1\}$ and $t_{i}\in\mathbb{N}\cup\{0\}$
$\forall i\in\{1,2,...,n\}$ and $n$ random real numbers \{$\theta_{1},\theta_{2},...,\theta_{n}$\}
such that $\theta_{i}\in[0,2\pi]$. He prepares $n$ quantum walk
states $[U(\theta_{i})]^{t_{i}}|x_{i}\rangle|c_{i}\rangle=U^{t_{i}}|x_{i}\rangle|c_{i}\rangle$
~~$\forall i\in\{1,2,...,n\}$ and sends these states to Alice. 
\item On receiving the walk states, Alice randomly chooses $n/2$ of them
for checking eavesdropping and classically sends their corresponding
coordinates $i$ to Charlie. Charlie classically sends to Alice the
corresponding values of $t_{i},x_{i},c_{i},$ and $\theta_{i}$. Alice
applies the operation $U^{-t_{i}}$ on those states and measures them
and checks the measurement result with the value of $x_{i}$ and $c_{i}$.
If the error is within a tolerable limit, Alice continues to step
3. Otherwise, the protocol is aborted and they restart the protocol
from the beginning. 
\item Out of the remaining $n/2$ walk states, Alice chooses $n/4$ of them
for encoding the message. On each of those $n/4$ states, Alice encodes
a part of her message $a_{i}$ by applying the translation operator
$T(a_{i})$. She does nothing to the other $n/4$ states (let us call
them decoy states). She then chooses a random integer $k$ and coin
parameter $\theta_{r}$, applies $[U(\theta_{r})]^{k}$ on all the
$n/2$ states and sends the states to Bob. 
\item Once Bob confirms the receiving of the states, Alice publicly announces
the values of $\theta_{r}$ and $k$ and the coordinates of the decoy
states. Charlie, upon receiving the announcement, sends the $t_{i},x_{i},c_{i},$
and $\theta_{i}$ values of the decoy states to Bob. Bob
then applies the corresponding operator $U^{-t_{i}}[U(\theta_{r})]^{-k}$
on the decoy states and checks for the presence of Eve just like how
Alice does it in step 2. 
\item Meanwhile, Bob encodes his message $b_{i}$ on the remaining message
states by applying the translation operator $T(b_{i})$. Once he confirms
the absence of eavesdropping, Charlie sends the $t_{i},x_{i},\theta_{i}$
and $c_{i}$ values of the message states to Bob. Bob
applies the operator $U^{-t_{i}}[U(\theta_{r})]^{-k}$ on the message
states, measures them and publicly announces the measurement results
$a_{i}+b_{i}$. Alice and Bob subtract $a_{i}$ and $b_{i}$ respectively
from their results to obtain each others' messages. 
\end{enumerate}

\noindent
Compared to LM05/DL04 protocols (see Appendix) which can transfer one 
bit per quantum state,  the discrete-time quantum walk protocol on a coin and position Hilbert space 
presented above can transfer more number of bits per quantum state allowing for faster 
transmission of message. In addition to this, the security advantage is discussed below.


\section{Security\label{sec:Security}}


In this section, we analyse the security of our protocol against various
attacks, namely the intercept-resend attack, the denial of service
attack, man-in-the-middle attack, and the attack by an untrusted Charlie.


\subsection{Intercept-and-resend attack\label{subsec:Intercept-and-Resend-Attack}}


In this attack, Eve intercepts the quantum channel and tries to extract
information from the incoming state by measuring it. Then, she re-prepares
the appropriate state (based on the information she receives) and
sends it to the receiver. Our protocols are robust against this attack.
This is because the discrete-time quantum walk states are usually
superposition states where the position and the coin Hilbert spaces
are usually entangled. Hence, Eve can't determine the incoming state
by measurement alone. Instead of directly measuring the state, Eve
can apply $U^{-t_{i}}$ and then measure the state. But this attack
also cannot be performed by Eve because the value of $t_{i}$ will
be only known to Alice at the time of attack. If Eve attempts to perform
this attack, she will raise the error during the eavesdropping checking
of the control mode states, and hence will be caught.


\subsubsection{Mutual Information between Alice and Eve\label{subsec:Mutual-Information-between}}


In practical scenarios, Alice can choose her parameters $t_{i},x_{i},c_{i},$
and $\theta_{i}$ only from a finite set or a finite range of values.
Hence, the amount of mutual information $I_{AE}$ gained between Alice
and Eve during the intercept-resend attack is dependent upon the size
of these sets and ranges. The higher the mutual information, the more
will be known by Eve about the state sent by Alice, thus making the
protocol less secure. Let us consider a practical scenario where Alice
can choose: 
\begin{itemize}
\item $t_{i}$ from the set $T$ containing $n(T)$ integers (from $0$
to $n(T)-1$) 
\item $x_{i}$ from the set $X=$\{$0,1,2,...,N-1$\} (set of $N$ values),
$N$ being the dimension of the position space 
\item $c_{i}$ from the set $C=$\{$0,1$\} (set of 2 values) 
\item $\theta_{i}$ from the range $R_{\theta}=[\theta_{min},\theta_{max}]$ 
\end{itemize}
Let us say, that for a particular round of transmission, Alice chooses
the values $t_{A}\in T$, $x_{A}\in X$, $c_{A}\in C$, and $\theta_{A}\in R_{\theta}$
and prepares the state $|\psi_{A}\rangle=[U(\theta_{A})]^{t_{A}}|x_{A}\rangle|c_{A}\rangle$.
Now Eve can perform the intercept-resend attack in two ways, 
\begin{enumerate}
\item directly measure the incoming state to obtain the position and coin
values $x_{E}$ and $c_{E}$ respectively (Let us call this strategy
IR1) , or 
\item randomly choose the values $t_{E}\in T$, $x_{E}\in X$, $c_{E}\in C$,
and $\theta_{E}\in R_{\theta}$ and perform the operation $[U(\theta_{E})]^{-t_{E}}|\psi_{A}\rangle$
and then measure the position and coin values of the resulting state
in order to obtain the values $x_{E}$ and $c_{E}$ respectively (let
us call this strategy IR2). 
\end{enumerate}
Let us now examine IR2. We can consider $t_{A},\,x_{A},\,c_{A},\,t_{E},\,x_{E},\,c_{E},\,\theta_{A},$
and $\theta_{E}$ as uniformly distributed random variables, where
$t_{A},\,x_{A},\,c_{A},\,t_{E},\,x_{E},\,c_{E}$ are discrete and
$\theta_{A}$ and $\theta_{E}$ are continuous. Now, for IR2, the
mutual information $I_{AE_{2}}$ between Alice and Eve is given by, 
\begin{center}
\begin{multline}
I_{AE_{2}}=\underset{t_{E}\in T}{\sum}\underset{x_{E}\in X}{\sum}\underset{c_{E}\in C}{\sum}\underset{t_{A}\in T}{\sum}\underset{x_{A}\in X}{\sum}\underset{c_{A}\in C}{\sum}\underset{\theta_{A}=\theta_{min}}{\overset{\theta_{max}}{\int}}\underset{\theta_{E}=\theta_{min}}{\overset{\theta_{max}}{\int}}\\
p(t_{A},x_{A},c_{A},t_{E},x_{E},c_{E},\theta_{A},\theta_{E})log_{2}\frac{p(t_{A},x_{A},c_{A},t_{E},x_{E},c_{E},\theta_{A},\theta_{E})}{p(t_{A})p(x_{A})p(c_{A})p(t_{E})p(x_{E})p(c_{E})p(\theta_{A})p(\theta_{E})}d\theta_{A}d\theta_{E},
\end{multline}
\par\end{center}

\noindent where $p(a_{1},a_{2},...,a_{n})$ is the joint probability
distribution-mass function of the random variables $a_{1},a_{2},...,a_{n}$
where $a_{i}\in\{t_{A},x_{A},c_{A},t_{E},x_{E},c_{E},\theta_{A},\theta_{E}\}$.

\noindent For IR1, the mutual information $I_{AE_{1}}$ between Alice
and Eve is given by, 
\begin{center}
\begin{multline}
I_{AE_{1}}=\underset{x_{E}\in X}{\sum}\underset{c_{E}\in C}{\sum}\underset{t_{A}\in T}{\sum}\underset{x_{A}\in X}{\sum}\underset{c_{A}\in C}{\sum}\underset{\theta_{A}=\theta_{min}}{\overset{\theta_{max}}{\int}}p(t_{A},x_{A},c_{A},x_{E},c_{E},\theta_{A})log_{2}\frac{p(t_{A},x_{A},c_{A},x_{E},c_{E},\theta_{A})}{p(t_{A})p(x_{A})p(c_{A})p(x_{E})p(c_{E})p(\theta_{A})}d\theta_{A}.
\end{multline}
\par\end{center}

\noindent The above formulas of $I_{AE_{1}}$ and \emph{$I_{AE_{2}}$}
contain 1 and 2 integrals respectively. Due to lack of access to good
computing power to calculate $I_{AE_{1}}$ and \emph{$I_{AE_{2}}$},
we modify the protocol for the purpose of analysis of this attack,
by keeping all the coin parameters, including $\theta$ constant and
publicly known throughout the protocol, thus reducing the number of
secret parameters and avoiding the integrals. Now, the revised formulas
for $I_{AE_{1}}$ and $I_{AE_{2}}$ will be 
\begin{center}
\begin{equation}
I_{AE_{2}}=\underset{t_{E}\in T}{\sum}\underset{x_{E}\in X}{\sum}\underset{c_{E}\in C}{\sum}\underset{t_{A}\in T}{\sum}\underset{x_{A}\in X}{\sum}\underset{c_{A}\in C}{\sum}p(t_{A},x_{A},c_{A},t_{E},x_{E},c_{E})log_{2}\frac{p(t_{A},x_{A},c_{A},t_{E},x_{E},c_{E})}{p(t_{A})p(x_{A})p(c_{A})p(t_{E})p(x_{E})p(c_{E})}
\end{equation}
\par\end{center}

\noindent and 
\begin{center}
\begin{equation}
I_{AE_{1}}=\underset{x_{E}\in X}{\sum}\underset{c_{E}\in C}{\sum}\underset{t_{A}\in T}{\sum}\underset{x_{A}\in X}{\sum}\underset{c_{A}\in C}{\sum}p(t_{A},x_{A},c_{A},x_{E},c_{E})log_{2}\frac{p(t_{A},x_{A},c_{A},x_{E},c_{E})}{p(t_{A})p(x_{A})p(c_{A})p(x_{E})p(c_{E})}
\end{equation}
\par\end{center}

\noindent where 
\begin{center}
\begin{equation}
p(t_{A},x_{A},c_{A},t_{E},x_{E},c_{E})=\frac{`1}{2N[n(T)]^{2}}(\langle x_{E}|\langle c_{E}|U{}^{-t_{E}}U{}^{t_{A}}|x_{A}\rangle|c_{A}\rangle)^{2},
\end{equation}
\par\end{center}

\begin{center}
\begin{equation}
p(t_{A},x_{A},c_{A},x_{E},c_{E})=\frac{`1}{2N[n(T)]}(\langle x_{E}|\langle c_{E}|U{}^{t_{A}}|x_{A}\rangle|c_{A}\rangle)^{2}
\end{equation}
\par\end{center}

\noindent and 
\begin{center}
\begin{equation}
p(a_{i})=\underset{a_{1,},a_{2},...,a_{i-1},a_{i+1},...,a_{n}}{\sum}p(a_{1},a_{2},...,a_{n})
\end{equation}
\par\end{center}

\noindent where $a_{j}\in\{t_{A},x_{A},c_{A},t_{E},x_{E},c_{E}\}$
and $U=U(\theta)$ where $\theta$ is the publicly known coin parameter
throughout the protocol. We can see that $I_{AE_{1}}$ and \emph{$I_{AE_{2}}$}
are a function of $n(T)$ and $N$, and also depend on the fixed coin
parameter $\theta$. 

\begin{center}
\begin{figure}[h]
\begin{centering}
\includegraphics[scale=0.6]{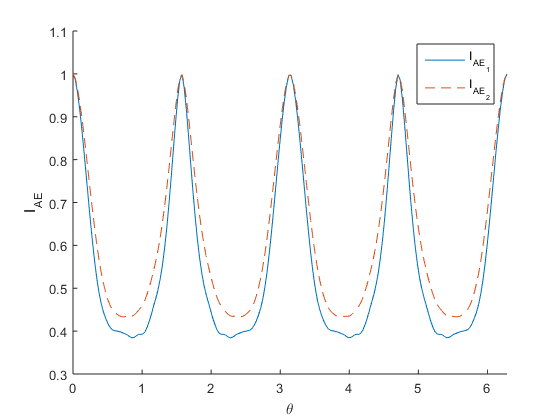}\includegraphics[scale=0.6]{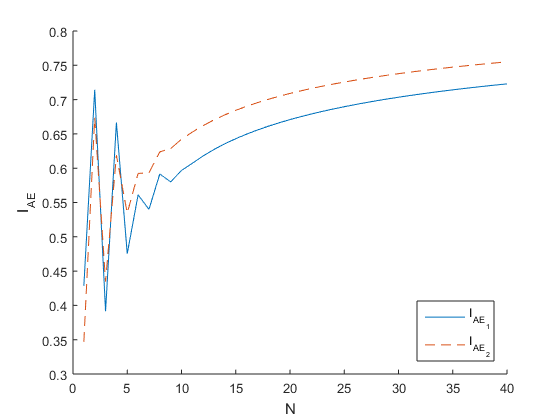} 
\par\end{centering}
\caption{(a) Mutual Information $I_{AE}$ vs coin parameter $\theta$ for $N=3,n(T)=7$.
(b) Mutual Information $I_{AE}$ vs the cycle length $N$ for $n(T)=7,\theta=\frac{\pi}{4}$.
The variation of $I_{AE}$ with $\theta$ is periodic with period
$\frac{\pi}{2}$ with the peaks at even multiples of $\frac{\pi}{4}$
and minimum at odd multiples of $\frac{\pi}{4}$. The variation of
$I_{AE}$ with $N$ is fluctuating in the beginning, but later steadily
increases. For both the plots, the coin parameters $\zeta$ and $\xi$
were set to a fixed value $\frac{\pi}{4}$.}
\label{fig-3} 
\end{figure}
\par\end{center}


In Fig. \ref{fig-3}(a), we can see that $I_{AE}$ is at its lowest
when $\theta$ is an odd multiple of $\frac{\pi}{4}$ and is at its
highest ($I_{AE}=1$) when $\theta$ is an even multiple of $\frac{\pi}{4}$.
Hence, for $\theta$ equal to even multiples of $\frac{\pi}{4}$,
the security of the protocol will be compromised. This is consistent
with the discrete-time quantum walk dynamics, for $\theta$ being
even multiples of $\frac{\pi}{4}$, the walk will either be localized
around the origin or will be ballistic without being in superposition
of more than two position space at a time \cite{meyer1996quantum,SU2}.
We can infer that the degree of spread of the walker in position space
gives an enhanced security to the protocol. In Fig. \ref{fig-3}(b),
we can see that for odd $N$, $I_{AE}$ increases with increase in
$N$, whereas for even $N$, $I_{AE}$ initially decreases with $N$,
but then increases. In Fig. \ref{fig-5}, we see that $I_{AE}$ decreases
with $n(T)$ and its value is greater for even $N$ than for odd $N$.
In fact, for odd $N$, the $I_{AE}$ drops much below
0.5 (which is the $I_{AE}$ value for the LM05/DL04 protocol (see
appendix)) for large $n(T)$, and in fact is less than 0.25 for $n(T)>25$.
From Fig. \ref{fig-5}(b) and Fig. \ref{fig-3}, we can see that $I_{AE_{2}}>I_{AE_{1}}$,
implying that that IR2 is a better strategy for Eve than IR1 for odd
$N$. This shows that, for an odd, low value of $N$, and a high value
of $n(T)$, and $\theta$ being an odd multiple of $\frac{\pi}{4}$,
our discrete-time quantum walk protocols are more secure against the
intercept-resend attack than the LM05/DL04 protocol (whose $I_{AE}=0.5$),
even with the modification that the coin parameters remain constant
and publicly known throughout the protocol. 
\begin{center}
\begin{figure}[h]
\begin{centering}
\includegraphics[scale=0.6]{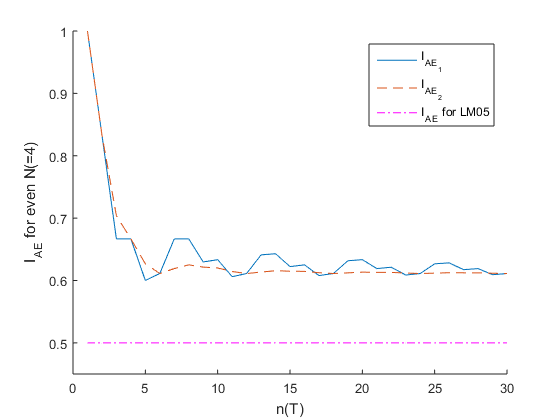}\includegraphics[scale=0.6]{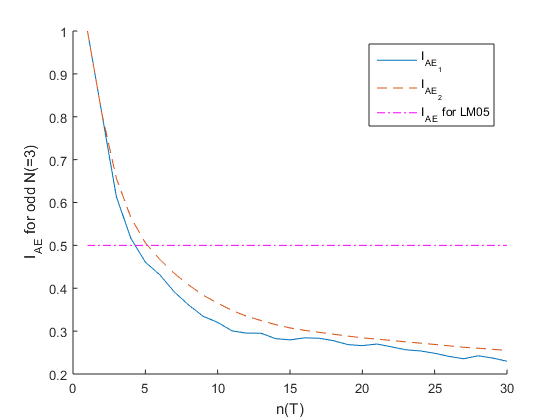} 
\par\end{centering}
\caption{Mutual Information $I_{AE}$ vs $n(T)$ for (a) $N=4$ and (b) $N=3$.
The dash-dotted line (magenta coloured in online
and colour-printed versions) represents the $I_{AE}$ for one channel
of the LM05/DL04 protocol, which is the same as the $I_{AE}$ for
the BB84 protocol. In both the figures, $I_{AE}$ decreases with
$n(T)$, thus increasing security of the protocol with increase in
$n(T)$. For both the plots, the coin parameters $\theta,\,\zeta$
and $\xi$ were set to a fixed value $\frac{\pi}{4}$.}
\label{fig-5} 
\end{figure}
\par\end{center}


\subsection{Denial of Service attack\label{subsec:Denial-of-Service}}


Instead of trying to extract information from the incoming state,
Eve can rather perform a denial-of-service attack i.e., she can just
stop the incoming state from going forward and can instead prepare
and send a random discrete-time quantum walk state. This attack also
cannot be performed by Eve because if she does so, she introduces
an added error and noise into the channel, and hence the eavesdropping
checking performed by the sender and the receiver at each quantum
channel will detect Eve.


\subsection{Man-in-the-middle attack\label{subsec:Man-in-the-middle-attack}}


Let's consider the QSDC protocol. In this attack, Eve initially puts
the incoming state from Alice into her quantum memory. Then, she sends
her own walk state to Bob. Bob, assuming that Alice may have sent
this state, encodes his message on this state and sends it back to
Alice. Eve intercepts that channel also and reads the message. She
then encodes the message onto Alice's state which she had earlier
stored in her quantum memory and sends it back to Alice, thus being
able to read the message. Eve can perform a similar kind of attack
in the CQD protocol to obtain the message of one of the two communicating
parties. In both cases of this attack, Eve will be detected by the
communicating parties during eavesdropping checking. Hence both our
protocols are unconditionally secure against this attack.


\subsection{Attack by an untrusted Charlie\label{subsec:Attack-by-an}}


Let us consider the QDC protocol. In this attack, Charlie intercepts
the Alice-Bob channel, applies $U^{-t_{i}}$ on the incoming state,
and obtains Alice's message by measuring the state. Then, he re-prepares
the state and sends it to Bob. Then when Bob encodes his message $b_{i}$
and announces the value $a_{i}+b_{i}$, Charlie can then get Bob's
message as well. But our QDC protocol is robust against this attack
because as Alice applies an additional $[U(\theta_{r})]^{k}$ to the
states, Charlie will not know the value of $\theta_{r}$ or $k$ and
hence he cannot apply $[U(\theta_{r})]^{-k}$ to retrieve the state.


\section{Conclusion\label{sec:Conclusion}}


The unique features of a discrete-time quantum walk such as spreading
the quantum state in the superposition of position space and entanglement
generation between the position and the coin states have an immense
unexplored potential for quantum security for communication and cryptographic
protocols. In this work, we have explored its potential for providing
cryptographic security by proposing two new protocols, a one-way two-party
Quantum Secure Direct Communication (QSDC) protocol and a two-way
three-party Controlled Quantum Dialogue (CQD) protocol. We have shown
that the proposed protocols are unconditionally secure against various
attacks, such as the intercept-resend attack, the denial of service
attack, and the man-in-the-middle attack. The CQD protocol, in particular,
is shown to be secure against an attack by an untrusted Charlie. Also,
for the intercept-resend attack, the mutual information gained between
Alice and Eve is shown to be much lower for the proposed protocols
as compared to the qubit based protocols such as the LM05/DL04 protocol
\cite{key-6,DL04}, thus making the proposed protocols more secure
than the LM05/DL04 against this attack. Also, unlike the qubit based protocols
which transfer just one bit per state, the proposed protocols can
transfer multiple bits per state \cite{SC19}, which can possibly
lead to advantages such as the faster transmission of messages and
a lower requirement of resources (both subject to practical/experimental
conditions). These direct communication schemes could potentially
lead to secure feasible solutions for many social and economic problems
such as the socialist millionaire problem \cite{key-17}, quantum
E-commerce \cite{key-18}, quantum voting \cite{key-19} and the work
towards finding these potential solutions are to be attempted in the
future.


\begin{acknowledgments}
SS and CMC would like to thank Department of Science and Technology,
Government of India for the Ramanujan Fellowship grant No.:SB/S2/RJN-192/2014.
We also acknowledge the support from Interdisciplinary Cyber Physical
Systems (ICPS) programme of the Department of Science and Technology,
India, Grant No.:DST/ICPS/QuST/Theme-1/2019/1. 
\end{acknowledgments}



\section*{Appendix}


\subsection*{LM05/DL04 Protocol}


This qubit based protocol was introduced in \cite{key-6,DL04}. In
this protocol, the encoding rules for the message sender are as follows:

To encode the bit 0, do nothing to the incoming qubit.

To encode the bit 1, apply the operator $iY=ZX$ on the incoming qubit.
The transformations are as follows: 
\begin{center}
$iY|0\rangle=-|1\rangle$ 
\par\end{center}

\begin{center}
$iY|1\rangle=|0\rangle$ 
\par\end{center}

\begin{center}
$iY|\pm\rangle=\pm|\mp\rangle$ 
\par\end{center}

The protocol is as follows: 
\begin{enumerate}
\item Alice chooses $n$ random qubits from the set \{$|0\rangle,|1\rangle,|+\rangle,|-\rangle$\}
and sends them to Bob. 
\item Out of these $n$ qubits received from Alice, Bob randomly chooses
$n/2$ of them and classically sends their coordinates to Alice. 
\item Alice publicly announces the states of the $n/2$ qubits which Bob
chose in step 2. Bob measures each of the $n/2$ qubits in their corresponding
bases and checks for eavesdropping. If the error is within a tolerable
limit, then the protocol continues to step 4. Else, the protocol is
discarded and they start all over again. 
\item Among the remaining $n/2$ qubits, Bob randomly chooses $n/4$ of
them and encodes the message in them according to the encoding rules
above and does nothing to the remaining $n/4$ qubits. He sends all
these $n/2$ qubits back to Alice. 
\item After Alice confirms receiving the $n/2$ qubits, Bob sends the coordinates
of the qubits on which he didn't encode the message. Alice uses these
qubits to check for eavesdropping just like how Bob does it in step
3. 
\item After confirming no eavesdropping, Alice measures the remaining qubits
in their respective bases to obtain the message sent by Bob. 
\end{enumerate}

\subsection*{Mutual Information}

Let us take two random variables, say $x$ and $y$. The mutual information
$I_{XY}$ between two random variables $x$ and $y$ is the decrease
in uncertainty of one random variable when the value of the other
random variable is observed, measured or determined. If $x$ and $y$
are discrete, the formula for $I_{XY}$ is given by \cite{key-20}

\begin{equation}
I_{XY}=\underset{x}{\sum}\underset{y}{\sum}p(x,y)log_{2}\frac{p(x,y)}{p(x)p(y)}
\end{equation}

\noindent where $p(x,y)$ is the joint probability mass function and
$p(x)$ and $p(y)$ are the individual probability mass functions.

\noindent If $x$ and $y$ are continuous, then the formula for $I_{XY}$
is given by

\begin{equation}
I_{XY}=\underset{x}{\int}\underset{y}{\int}p(x,y)log_{2}\frac{p(x,y)}{p(x)p(y)}dxdy
\end{equation}

\noindent Where $p(x,y)$ is the joint probability density function
and $p(x)$ and $p(y)$ are the individual probability density functions.

\noindent There can also be a case where one of the random variables
is discrete and the other is continuous. For example, if $x$ is discrete
and $y$ is continuous, then the formula for $I_{XY}$ becomes

\begin{equation}
I_{XY}=\underset{x}{\sum}\underset{y}{\int}p(x,y)log_{2}\frac{p(x,y)}{p(x)p(y)}dy
\end{equation}

\noindent where $p(x)$ is the probability mass function of $x$,
$p(y)$ is the probability density function of $y$ and $p(x,y)$
is a function that is a probability density-mass function that is
discrete in $x$ and continuous in $y$.

\noindent This concept of mutual information can also be generalized
to $r=mn>2$ random variables $\{x_{1},x_{2},...,x_{m}\}$ and $\{y_{1},y_{2},...,y_{n}\}$
where $x_{i}$ are discrete and $y_{i}$ are continuous. The generalised
mutual information $I_{mutual}$ is given by \cite{key-20}

\begin{equation}
I_{mutual}=\underset{x_{1},x_{2},...,x_{n}}{\sum}\underset{y_{1},...,y_{n}}{\int}p(x_{1},x_{2},...,x_{m},y_{1},y_{2},...,y_{n})log_{2}\frac{p(x_{1},x_{2},...,x_{m},y_{1},y_{2},...,y_{n})}{p(x_{1})p(x_{2})...p(x_{m})p(y_{1})p(y_{2})...p(y_{n})}dy_{1}dy_{2}...dy_{n}.
\end{equation}


\subsubsection*{Mutual Information for the intercept-resend attack for the LM05/DL04
protocol:}


Let us consider the first transmission from Alice to Bob. In this
transmission, Alice first selects either of the four states and prepares
them and sends them to Bob. Eve intercepts this channel before the
state reaches Bob and randomly chooses a basis for each incoming state
and measures the state in that basis. Let $a,e\in\{0,1,+,-\}$. Let
the probability of Alice sending the qubit $a$ and Eve receiving
the qubit $e$ be $p(a,e)$. For example, the probability $p(0,0)$
is

\begin{equation}
p(0,0)=\overset{probability\,of\,Alice\,choosing\,0}{\frac{1}{4}}\times\overset{probability\,of\,Eve\,choosing\,the\,computational\,Z\,basis}{\frac{1}{2}}\times\overset{probability\,of\,Eve\,getting\,0}{1}=\frac{1}{8}.
\end{equation}

\noindent Similarly,

\begin{equation}
p(0,1)=\frac{1}{4}\times\frac{1}{2}\times0=0,
\end{equation}

\begin{equation}
p(0,+)=\frac{1}{4}\times\frac{1}{2}\times\frac{1}{2}=\frac{1}{16},
\end{equation}

\begin{equation}
p(0,-)=\frac{1}{4}\times\frac{1}{2}\times\frac{1}{2}=\frac{1}{16},
\end{equation}

\noindent and similar probabilities for $p(1,e),p(+,e)$, and $p(-,e),$
where $e\in\{0,1,+,-\}$.

\noindent Hence, the mutual information $I_{AE}$ for the LM05/DL04
protocol is given by

\begin{equation}
I_{AE}=\underset{a}{\sum}\underset{e}{\sum}p(a,e)log_{2}\frac{p(a,e)}{p(a)p(e)}
\end{equation}

\begin{center}
$=4(\frac{1}{8}log_{2}\frac{\frac{1}{8}}{\frac{1}{16}}+\frac{1}{16}log_{2}\frac{\frac{1}{16}}{\frac{1}{16}}+\frac{1}{16}log_{2}\frac{\frac{1}{16}}{\frac{1}{16}})=0.5$
. 
\par\end{center}

\noindent (We can see that for all $a$ and $e$, $p(a)=p(e)=\frac{1}{4}$.
Hence, $p(a)p(e)=\frac{1}{16}$).



\end{document}